\documentclass[twocolumn,twoside,aps,pra,showpacs,superscriptaddress,floatfix]{revtex4}
\usepackage{epsf,epsfig,amsfonts,amsmath,amssymb,graphicx,times}
\usepackage[utf8]{inputenc}
\usepackage[OT1]{fontenc}
\usepackage{mathtools,color,stmaryrd,hyperref}

\begin{document}
\title{Unidimensional continuous-variable quantum key distribution using squeezed states}
\author{Vladyslav C. Usenko}
\email{usenko@optics.upol.cz}
\affiliation{Department of Optics, Palacký University, 17. listopadu 50,  
  772~07 Olomouc, Czech Republic}
\affiliation{Bogolyubov Institute for Theoretical Physics of National Academy of Sciences,
Metrolohichna st. 14-b, 03680, Kiev, Ukraine}
\begin{abstract}
The possibility of using squeezed states in the recently suggested unidimensional continuous-variable quantum key distribution based on a single quadrature modulation is addressed. It is shown that squeezing of the signal states expands the physicality bounds of the effective entangled state shared between the trusted parties due to the antisqueezing noise in the unmodulated quadrature. Modulation of the antisqueezed quadrature, on the other hand, effectively shrinks the physicality bounds due to the squeezing in the unmodulated quadrature and also provides noise on the reference side of the protocol, thus limiting the possibility of eavesdropping in noisy channels. This strategy is practical for low-loss (i.e., short-distance) channels, especially if direct reconciliation scheme is applied.
\end{abstract}
\pacs{03.67.Hk, 03.67.Dd}
\maketitle

\section{Introduction}

Quantum key distribution (QKD) \cite{Scarani2009} is the practical application of quantum information science, which is aimed at the development of methods (protocols) for the distribution of secure keys such that the security of a key is provided by the laws of quantum physics. The key can be used later in classical one-time pad cryptography, thus providing the complete postquantum solution for secure communication resilient against foreseen effective quantum computing. After starting with qubit-based discrete-variable protocols (see \cite{Gisin2002} for review) QKD was recently extended to continuous-variable (CV) \cite{Braunstein2005} protocols (see \cite{Diamanti2015} for review) which are aimed at providing higher key rates and simpler implementation compared to their discrete-variable counterparts. The first ideas in the field of CV QKD were based on discrete modulation and decoding of coherent and quadrature-squeezed \cite{Ralph1999} states of light as well as photon-number squeezed states \cite{Funk2002,Usenko2007,Usenko2010}, but had limited security proofs. It was an important step in the development of CV QKD when the use of Gaussian modulation \cite{Weedbrook2012a} was suggested \cite{Cerf2001} for quadrature-squeezed states \cite{Lvovsky2014} and later shown to be applicable for coherent states as well \cite{Grosshans2002}. It was shown that Gaussian protocols using squeezed \cite{Cerf2001,Garcia2009,Usenko2011,Madsen2012} and coherent \cite{Grosshans2002,Grosshans2003,Weedbrook2004,Lance2005,Lodewyck2007,Pirandola2008,Jouguet2013,Huang2015,Huang2016} states are secure against collective attacks and can tolerate, in principle, any level of channel attenuation if reverse information reconciliation is used \cite{Grosshans2003}. In addition, a family of measurement-device-independent CV QKD protocols was developed and tested using coherent states of light \cite{Pirandola2013,Pirandola2015a}. Importantly, security of CV QKD against collective attacks implies security against general attacks in the asymptotic limit \cite{Renner2009,Tomamichel2012,Leverrier2013} of an infinite number of data as well as, under certain constraints, in the finite-size regime \cite{Furrer2012,Furrer2014,Leverrier2015,Leverrier2017}, when the number of data is finite.  

Security analysis of Gaussian CV QKD protocols against collective attacks is based on the extremality of Gaussian states \cite{Wolf2006} and subsequent optimality of Gaussian collective attacks \cite{Navascues2006,Garcia2006}. This enables security analysis based on the covariance matrix formalism, which is sufficient for characterization of Gaussian states, and imposes that trusted parties perform the channel estimation and are able to derive the covariance matrix of an entangled state effectively shared between them \cite{Grosshans2003a}. In order to know the channel properties the trusted parties perform modulation and measurement of both the complementary quadratures and then optimally switch between channel estimation and key distribution \cite{Ruppert2014}. Therefore, both the amplitude and the phase modulators must be employed by a trusted sender party in order to apply Gaussian modulation of amplitude and phase quadratures. Recently a simplified unidimensional (UD) CV QKD protocol \cite{Usenko2015} was suggested and experimentally tested \cite{Gehring2016} on the basis of coherent states of light in order to provide simpler implementation potentially based on a single (e.g., phase) quadrature modulation with no need to perform modulation in a complementary quadrature. It was shown that if the remote trusted party is able to estimate the variance of the unmodulated quadrature, an eavesdropper can be limited by the physicality bounds on the effective entangled state shared between the trusted parties, and security of the protocol can be accessed by using a pessimistic assumption on the correlation between the sender and the receiver in the unmodulated quadrature. 

It was previously shown that the use of squeezed states can make CV QKD more robust against imperfections, such as channel noise and limited postprocessing efficiency \cite{Garcia2009,Usenko2011,Madsen2012}; moreover, squeezing, if used optimally, can potentially eliminate information leakage from purely attenuating channels \cite{Jacobsen2018}. Moreover, the use of squeezed states in CV QKD becomes more and more feasible, in particular, with the development of compact on-chip squeezers \cite{Dutt2015,Masada2016}. Thus it is important to verify the effect of signal-state squeezing and identify its possible advantages in UD CV QKD.

In the current paper we generalize the result considering the use of quadrature-squeezed signal states in UD CV QKD. We show that the presence of antisqueezing noise makes the protocol worse compared to its coherent-state counterpart if the squeezed quadrature is modulated. On the other hand, we show, surprisingly, a positive effect arising from the modulation of the noisy antisqueezed quadrature, which is concerned with the fact that the unmodulated quadrature remains squeezed and therefore allows better tolerance of channel noise and losses if direct information reconciliation is used. We therefore suggest the effective UD CV QKD protocol for short-distance channels, which benefits from the reduced fluctuations of the unmodulated quadrature and the trusted excess noise present in the modulated quadrature. Thus we fill the gap in the analysis of UD CV QKD by studying squeezed-state protocols as well as suggest the improvement of UD CV QKD by using modulation of the antisqueezed quadrature, which increases the key rate, secure distance of the protocol, and robustness to noise of the UD CV QKD with direct reconciliation, thereby contributing to solution of the major current challenges in QKD \cite{Diamanti2016}. The paper is structured as follows: in Sec. \ref{sec:prot} we consider the generalized squeezed-state UD CV QKD protocol; in Sec. \ref{sec:sec} we study the security and physicality bounds in the general phase-insensitive channels; in Sec. \ref{sec:sym} we consider the typical case of phase-insensitive Gaussian channels, and we devise analytical expressions for lower bounds on the secure key rate in the limit of strong modulation and compare the performance of the protocols based on the modulation of coherent and squeezed states; and in Sec. \ref{sec:sum} we give our Summary and Conclusions.

\section{Squeezed-state based unidimensional CV QKD protocol}
\label{sec:prot}
\begin{figure}
\includegraphics[width=0.9\columnwidth]{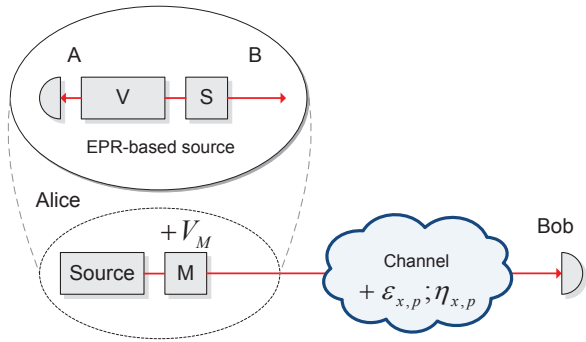}
\caption{Scheme of the squeezed-state UD CV QKD protocol. Alice prepares quadrature-squeezed states using, e.g., an optical parametric oscillator, and then modulates a state by
displacing it along the modulated quadrature using the modulator M so that the modulation variance is $V_M$. The states travel through an untrusted, generally phase-sensitive channel (with transmittance values $\eta_x$ and $\eta_p$ and excess noise values $\epsilon_x$ and $\epsilon_p$ in the $x$- and $p$-quadratures respectively) to a remote party, Bob, who performs homodyne measurement of the modulated quadrature. Inset: The equivalent entanglement-based scheme using a two-mode squeezed vacuum source: mode A is measured by Alice using a homodyne detector, and mode B is squeezed on the squeezer S and sent to a channel.
\label{scheme}}
\end{figure}

\begin{figure}
\includegraphics[width=0.9\columnwidth]{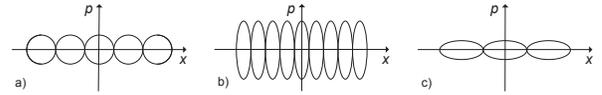}
\caption{Modulation schemes for unidimensional protocols: (a) using coherent states \cite{Usenko2015}, (b) using $x$-quadrature squeezed states and modulation in the squeezed quadrature, and (c) using $p$-quadrature squeezed states and modulation in the antisqueezed quadrature.
\label{mod_schemes}}
\end{figure}

We consider the protocol based on the preparation of quadrature-squeezed states \cite{Lvovsky2014} (e.g., using an optical parametric oscillator) and their subsequent Gaussian modulation in one of the quadratures. The scheme is depicted in Fig. \ref{scheme}. 
In the following with no loss of generality we assume that the signal states are squeezed or antisqueezed in $x$-quadrature. Therefore if the source generates pure $x$-quadrature-squeezed states characterized by the quadrature values $x_S$ and $p_S$, respectively, their variances are $Var(x_S)=V_S<1$ and $Var(p_S)=1/V_S>1$. Alternatively, the source can generate $p$-quadrature-squeezed states so that $Var(x_S)=V_S>1$ and $Var(p_S)=1/V_S<1$. The modulator then displaces $x$-quadrature, therefore performing modulation of the squeezed or antisqueezed quadrature, so that the modulated quadrature of the signal state sent to the channel in the case of $x$-quadrature modulation becomes $x_A=x_S+x_M$, where $x_M$ is the value of displacement, randomly picked from a zero-centered Gaussian distribution with variance $V_M$ and so $Var(x_A)=V_S+V_M$ and $p_A\equiv p_S$ since no modulation was performed in the $p$-quadrature. Therefore two modulation schemes are possible in the case of squeezed states: modulation in the squeezed quadrature and modulation in the antisqueezed quadrature. The modulation schemes are depicted in Fig. \ref{mod_schemes} along with the single-quadrature modulation of coherent states \cite{Usenko2015,Gehring2016} for comparison. 
The states then travel through an untrusted, generally phase-sensitive channel, which is characterized by transmittance values $\eta_x$ and $\eta_p$ and excess noise values $\epsilon_x$ and $\epsilon_p$ in the $x$- and $p$-quadratures, respectively. After the channel, the states are measured by Bob using a homodyne detector set to measure the $x$- or $p$-quadrature. Bob has to switch between the quadratures often enough in order to characterize the variance in both quadratures and estimate the correlation in the modulated quadrature, but in the asymptotic limit the fraction of $p$-quadrature measurements can be assumed to be negligible \cite{Usenko2015}. After a sufficient number of runs of the protocol, Alice and Bob analyze the security of the protocol and perform error correction and privacy amplification in order to distill the key using either the direct or reverse reconciliation scheme when Alice or Bob are, respectively, the references sides for the error correction algorithms. In the following section we estimate the security region of the squeezed-state UD protocol depending on the modulation scheme and reconciliation direction and compare it to the coherent-state-based UD CV QKD protocol.
\section{Security of the squeezed-state UD protocol}
\label{sec:sec}
Let us analyze the security of the protocol against the optimal Gaussian collective attacks, which, as mentioned above, implies security against general attacks in the asymptotic limit. To do so we follow the purification-based security analysis, where Eve is assumed to be able to purify (i.e., control) all the noise added in the untrusted quantum channel. Following the extension of the classical Csisz\'ar–K\"orner theorem \cite{Csiszar1978} to the quantum measurements, performed by Devetak and Winter \cite{Devetak2005}, the secure key can be distilled once the trusted parties Alice and Bob have the information advantage over the adversary Eve. Therefore, the protocol is secure once the lower bound on the key rate
\begin{equation}
\label{kr}
K_{DR}=\beta_{DR}I_{AB}-\chi_{AE}, K_{RR}=\beta_{RR}I_{AB}-\chi_{BE}
\end{equation}
is positive for direct (DR) or reverse (RR) reconciliation, i.e. when the mutual classical information between the trusted parties $I_{AB}$ exceeds the Holevo quantity \cite{Holevo2001} $\chi_{AE/BE}$. The latter upper bounds the information available to an eavesdropper on the key bits possessed by a reference-side trusted party respectively in the case of DR or RR. The mutual information between the trusted parties is scaled by the postprocessing efficiency $\beta_{DR/RR}\in(0,1)$, which depends on the effectiveness of the error correction algorithms for a given signal-to-noise ratio and is specific for a particular implementation of the protocol and direction of post-processing. In the current paper we aim to compare the ultimate performance of the protocols, therefore we set $\beta=1$; effects arising from the realistic finite-size regime shall not change the interplay between the protocols.

After the signal travels through the untrusted quantum channel, the trusted parties perform the estimation of the channel parameters, publicly revealing optimized fraction of the data \cite{Ruppert2014}. A Gaussian phase-sensitive channel acts as a linear map that transforms quadratures so that the output reads $\{x',p'\}=\sqrt{\eta_{\{x,p\}}}\{x,p\}+\{x,p\}_N+\sqrt{(1-\eta_{\{x,p\}})}\{x,p\}_0$, where $\eta_{\{x,p\}}$ are the channel transmittance values, and $\{x,p\}_N$ and $\{x,p\}_0$ are the excess and vacuum noise contributions, respectively, with $Var(\{x,p\}_N)=\epsilon_{\{x,p\}}$ and $Var(\{x,p\}_0)=1$ for the $x$- and $p$-quadratures. 

The classical mutual information $I_{AB}$ can be explicitly obtained from the variances and the correlations between the modulation data on the side of trusted sender (Alice) and measurement data on the side of trusted receiver (Bob) after the channel as $I_{AB}=\frac{1}{2}\log_2{V_A/(V_{A|B}}$, where $V_A=V_M$ is the variance of Alice's data, $V_{A|B}=V_A-C_{AB}^2/V_B$ is the conditional variance of Alice's data, $C_{AB}=\sqrt{\eta_x} V_M$ is the correlation between Alice's and Bob's data after the channel, and $V_B=\eta_x(V_S+V_M+\epsilon_x)+1-\eta_x$ is the variance of Bob's measured data after the channel. The mutual information then reads
\begin{equation}
\label{mutinf}
I_{AB}=\frac{1}{2}\log_2{\bigg[1+\frac{\eta_xV_M}{1+\eta_x(V_S+\epsilon_x-1)}\bigg]}
\end{equation}
and is the same for DR and RR protocols.

The calculation of the Holevo bound in either of the reconciliation scenarios is, however, more involved. In the case of Gaussian modulation, the Holevo bound is the difference $\chi_{AE}=S(E)-S(E|A)$ or $\chi_{BE}=S(E)-S(E|B)$ between the von Neumann entropy $S(E)$ of the state available to Eve for collective measurement and the von Neumann entropy of Eve's state conditioned by data on Alice's $S(E|A)$ or Bob's $S(E|B)$, respectively, for DR and RR. In the general case of channel noise being present it is assumed that Eve holds purification of the channel noise \cite{Navascues2006,Garcia2006} and then the equalities $S(E)=S(AB)$, $S(E|A)=S(B|A)$, and $S(E|B)=S(A|B)$ hold, where $S(AB)$ is the entropy of an initially pure state shared between the trusted parties through the noisy channel and $S(B|A)$, $S(A|B)$ are entropies of this state conditioned on the measurement results on Alice's or Bob's side in the DR and RR scenarios, respectively. Therefore, in order to assess the security of Gaussian CV QKD protocols in the case of collective attacks in noisy quantum channels, one needs to build an equivalent purification scheme, corresponding to the state preparation on Alice's side and state measurement on Bob's side. To do so for the UD squeezed-state CV QKD protocol, we start from the covariance matrix of a pure two-mode squeezed vacuum state with variance $V$, which purifies the Gaussian symmetrical modulation scheme \cite{Grosshans2003a}. In order to comply with the UD modulation of squeezed or antisqueezed states, we apply a squeezing operation on one of the modes with the squeezing parameter set to $-\log{VV_S}$. The resulting state is then described by the covariance matrix
\begin{multline}
\label{inputstate}
\gamma_{AB}=\\
 \begin{bmatrix}
	V & 0 & \sqrt{VV_S(V^2-1)} & 0 \\
	0 & V & 0 & \mathllap{-}\sqrt{\frac{V^2-1}{VV_S}} \\
	\sqrt{VV_S(V^2-1)} & 0 & V^2V_S & 0 \\
	0 & \mathllap{-}\sqrt{\frac{V^2-1}{VV_S}} & 0 & \frac{1}{V_S}
 \end{bmatrix}.
\end{multline}
It is easy to see that when Alice performs homodyne detection in the $x$-quadrature on mode $A$, she conditionally prepares the signal squeezed or antisqueezed state in mode B described by the diagonal covariance $\gamma_{B|x_A}=\gamma_B-\sigma_{AB}[x\gamma_Bx]^{MP}\sigma_{AB}=diag(V_S,1/V_S)$, where $\gamma_A$ and $\gamma_B$ are diagonal single-mode sub-matrices of (\ref{inputstate}) standing for modes A and B, respectively, $\sigma_{AB}$ is the off-diagonal correlation submatrix of (\ref{inputstate}), the diagonal matrix $x=diag(1,0)$ stands for homodyne detection in the $x$-quadrature, and $MP$ stands for the Moore-Penrose inverse of a matrix, applicable to singular matrices. On the other hand, the state of mode B, which is characterized by the diagonal single-mode covariance matrix $\gamma_B=diag(V^2V_S,1/V_S)$, corresponds to the modulated signal squeezed or antisqueezed state in the prepare-and-measure scheme once $V=\sqrt{1+V_M/V_S}$, then $\gamma_B=diag(V_S+V_M,1/V_S)$, which is exactly the same as the state of the signal mode, sent to the channel in the prepare-and-measure scheme. Therefore, the entanglement-based scheme, depicted in the inset in Fig. \ref{scheme}, is equivalent to the prepare-and-measure scheme based on squeezed or antisqueezed states with squeezed or antisqueezed variance $V_S$ modulated with modulation variance $V_M=V_S(V^2-1)$. 


Since the $p$-quadrature is not modulated, the correlation term in the unmodulated quadrature remains unknown to the trusted parties, similarly to the the coherent-state UD CV QKD \cite{Usenko2015}. Therefore, after the quantum channel, the covariance matrix of the initially pure state, (\ref{inputstate}), shared between Alice and Bob, in terms of the modulation variance $V_M$ reads
\begin{multline}
\label{outputstate}
\gamma_{AB}'=\\
 \begin{bmatrix}
    \sqrt{1+\frac{V_M}{V_S}} & 0 & \sqrt{\eta_xV_M}(1+\frac{V_M}{V_S})\mathrlap{^{\frac{1}{4}}} & 0 \\
    0 & \sqrt{1+\frac{V_M}{V_S}} & 0 & C_p \\
    \sqrt{\eta_xV_M}(1+\frac{V_M}{V_S})\mathrlap{^{\frac{1}{4}}} & 0 & V_x^B & 0 \\
    0 & C_p & 0 & V_p^B
 \end{bmatrix},
\end{multline}
where $C_P$ is the unknown correlation in the $p$-quadrature and $V_x^B=\eta_x(V_S+V_M+\epsilon_x)+1-\eta_x$. 

\begin{figure}[tb]
\includegraphics[width=0.9\columnwidth]{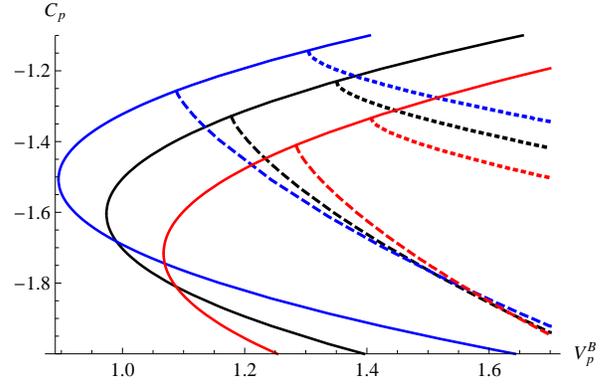}
\caption{Physicality (solid lines) and security within the physicality (dotted lines, DR; dashed lines, RR) regions of the UD protocol. Modulation variance $V_M=10$, channel transmittance in $x$ $\eta_x=0.9$, noise in $x$ $\epsilon_x=3\%$ SNR. Plots are given for coherent-state ($V_S=1$; middle, black lines), squeezed-state ($V_S=0.9$; lower, red lines), and antisqueezed-state ($V_S=1.1$; upper, blue lines) protocols. 
\label{cbounds}}
\end{figure}

The covariance matrices of the conditioned states after the signal propagation through the channel and after Alice's or Bob's measurements in the $x$-quadrature read, respectively,
\begin{equation}
\label{outputstateB}
\gamma_{B|x_A}'=
 \begin{bmatrix}
    \eta_x(V_S+\epsilon_x-1)+1 & 0 \\
    0 & V_p^B
 \end{bmatrix}
\end{equation}
and
\begin{equation}
\label{outputstateA}
\gamma_{A|x_B}'=
 \begin{bmatrix}
    \frac{\sqrt{1+\frac{V_M}{V_S}}[\eta_x(V_S+\epsilon_x-1)+1]}{V_x^B} & 0 \\
    0 & \sqrt{1+\frac{V_M}{V_S}}
 \end{bmatrix}.
\end{equation}
Now the Holevo bound can be assessed in either the DR or the RR scenario by calculating the von Neumann entropies of state (\ref{outputstateB}) or (\ref{outputstateA}), respectively, and subtracting them from the von Neumann entropy of the two-mode state, (\ref{outputstate}), which is done using the bosonic entropic function \cite{Serafini2005} of the symplectic eigenvalues of the respective covariance matrices \cite{Weedbrook2012a} (see \cite{Usenko2016} for details on Gaussian security analysis). The von Neumann entropy $S(AB)$ of the two-mode state, (\ref{outputstate}), then depends on the unknown correlation parameter $C_P$ in the unmodulated quadrature, which can, in principle, be set arbitrary by an eavesdropping attack in the untrusted channel. Nevertheless, the parameter can be bounded by the physicality considerations. Indeed, Eve's attack on the protocol should preserve the physicality of the state, effectively measured by Alice and Bob. In terms of the covariance matrix this is given by the constraint, following from the uncertainty principle,
\begin{equation}
\label{uncert}
\gamma_{AB}'+i\Omega \geq 0,
\end{equation}
where $\Omega$ is the symplectic form
\begin{equation}
\Omega=\bigoplus_{i=1}^{n}\omega\, , \quad \omega=
\left(\begin{array}{cc}
0&1\\
-1&0
\end{array}\right)\, , \label{symform}
\end{equation}
which imposes limitations on the possible values of $C_p$.
The physicality constraint in the general case can be represented by the parabolic equation on the $\{V_p^B,C_p\}$ plane,
\begin{equation}
\label{cpmax}
(C_p-C_0)^2\le\frac{V_M}{\sqrt{V_S(V_S+V_M)}}(1-\eta_xV_SV_0^B)(V_p^B-V_0^B)
\end{equation}
with vertex $(V_0^B,C_0)$, defined as
\begin{equation}
\label{vp0}
V_0^B=\frac{1}{1+\eta_x(V_S+\epsilon_x-1)}
\end{equation}
and
\begin{equation}
\label{c0}
C_0=-\frac{V_0^B\sqrt{\eta_xV_M}}{\big(\frac{V_M}{V_S}+1\big)^{\frac{1}{4}}}.
\end{equation}
The typical physicality regions are given in Fig. \ref{cbounds}. In addition, squeezing or antisqueezing of the signal and, respectively, antisqueezing or squeezing of the unmodulated quadrature also influence the security bounds of the protocol, given by the condition $K_{DR}=0$ or $K_{RR}=0$ for DR or RR, respectively. In the general case the security can be evaluated numerically and the typical bounds are given in Fig. \ref{cbounds} along the physicality bounds. 
%
It is evident from the physicality bounds plotted in Fig. \ref{cbounds}, that the use of squeezed or antisqueezed states as the signal carriers shifts the physicality region. Indeed, if the squeezed states are used, the region is shifted towards higher values of noise $V_p^B$ and expanded, because the antisqueezing noise present in the $p$-quadrature should result in above-shot-noise fluctuations in the $p$-quadrature and allows for a wider region of correlation term $C_p$ values than those for the coherent-state protocol. On the other hand, modulation in the antisqueezed quadrature shifts the physicality region to $V_p^B$ below shot noise, since the $p$-quadrature in this case is squeezed, and allows for a narrower region of correlation term values for given noise $V_p^B$. In the next section we consider the role of signal-state squeezing and antisqueezing in UD CV QKD protocols in the typical class of phase-insensitive Gaussian channels.
\section{Role of signal squeezing or antisqueezing in symmetrical channels}
\label{sec:sym}
In the previous section we have derived the general physicality and security bounds considering generally phase-sensitive channel, having different transmittance and excess noise in the $x$- and $p$-quadratures. However, in practice the quantum channels (fiber or free space) are typically inclined to the same transmittance and the same excess noise in both quadratures, thus being phase-insensitive (symmetric). In the current section we focus on the role of signal state squeezing and antisqueezing in UD CV QKD protocols over such channels. 

\begin{figure}[tb]
\includegraphics[width=0.9\columnwidth]{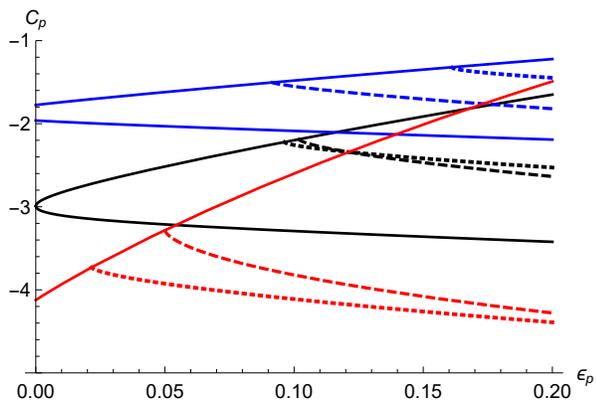}
\caption{Physicality (solid lines) and security within the physicality (dotted lines, DR; dashed lines, RR) regions of the UD protocol in channels with symmetric transmittance $\eta_x=\eta_p=0.9$ with respect to excess noise $\epsilon_p$. Modulation variance $V_M=10$, noise in $x$ $\epsilon_x=3\%$ SNR. Plots are given for coherent-state ($V_S=1$; middle, black lines), squeezed-state ($V_S=0.9$; lower, red lines) and antisqueezed-state ($V_S=1.1$; upper, blue lines) protocols. 
\label{cboundssym}}
\end{figure}

First, we assume that the channel transmittance is symmetrical, $\eta_x=\eta_p\equiv\eta$; then the structure of noise measured in the $p$-quadrature on Bob's side is $V_p^B=\eta(1/V_S+\epsilon_p)$. In Fig. \ref{cboundssym} we plot physicality and security bounds in this case similarly to the ones given in Fig. \ref{cbounds}. This allows us to compare the robustness of the UD protocols to channel noise.
It is evident from the plot that in the case of the squeezed-state protocol (lower, red lines) the security upon arbitrary $C_p$ is lost at lower excess noise for DR (dotted lines) than for RR (dashed lines) and, in both cases, at lower noise than for the coherent and antisqueezed protocols. On the other hand, the coherent-state protocol (middle, black lines) demonstrates almost the same tolerance to channel noise for RR and DR under a given transmittance and with the given modulation. Finally, the antisqueezed protocol (upper, blue lines) demonstrates a similar tolerance to channel noise as the coherent-state protocol for RR, but is more robust against channel noise in the case of DR. Indeed, the antisqueezed protocol allows for weaker noise-infusing attacks due to squeezing of the $p$-quadrature; on the other hand, it is known that the DR protocol is more robust against trusted preparation noise \cite{Weedbrook2010,Weedbrook2012,Usenko2016}. On the contrary, the squeezed-state UD protocol loses this advantage, allowing for broader attacks within the noisy antisqueezed $p$-quadrature, which is not compensated by having less noise in the modulated (squeezed) $x$-quadrature. 

\begin{figure}[h]
\begin{tabular}{ll}
\includegraphics[width=0.24\textwidth]{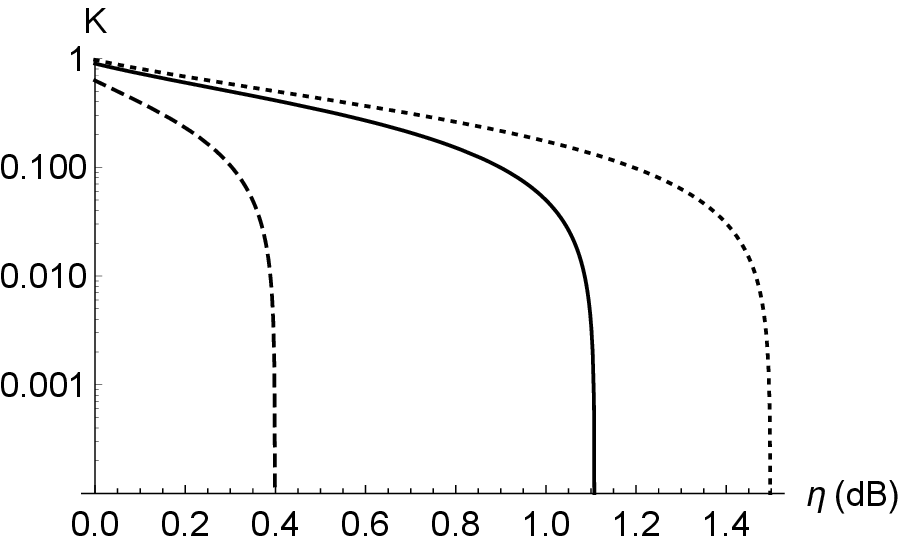}
\includegraphics[width=0.24\textwidth]{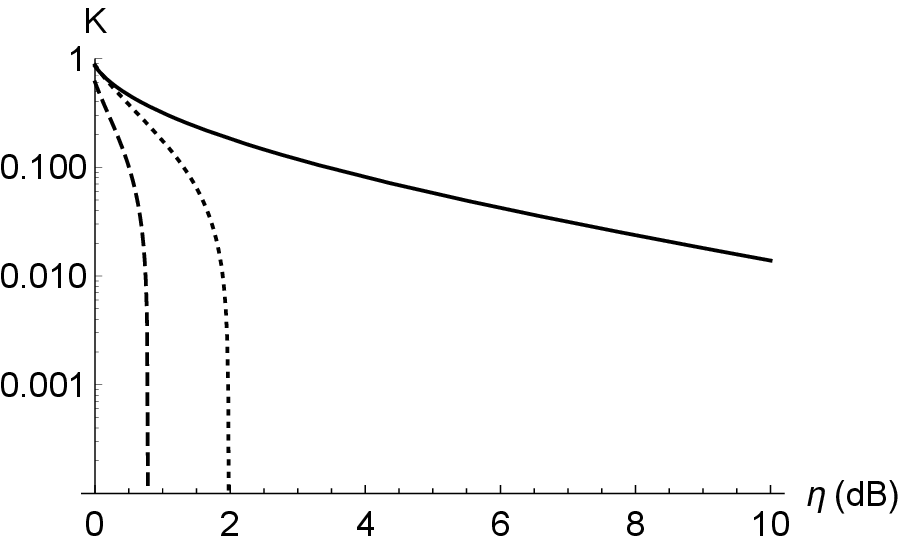}	
\end{tabular}
\caption{Secure key rate versus channel attenuation (on dB scale), secure against collective attacks for the DR (left) and RR (right) protocols, for the coherent-state (solid lines), squeezed-state ($V_S=0.5$; dashed lines), and antisqueezed-state ($V_S=2$; dotted lines) protocols. Channel noise $\epsilon=3\%$ SNU, modulation variance $V_M=100$. 
\label{KRvsLOSS}}
\end{figure}

The above given is confirmed in fully phase-insensitive (symmetrical) channels with the same transmittance as well as the same noise $\epsilon_x=\epsilon_p\equiv\epsilon$ in both quadratures. We first plot the lower bound on the key rate, (\ref{kr}), for the DR and RR squeezed-, coherent-, and antisqueezed-state protocols upon fixed channel excess noise in Fig. \ref{KRvsLOSS}.
It can be clearly seen, that upon RR the coherent-state protocol provides much better robustness against channel attenuation at given noise levels than the squeezed or antisqueezed protocol (therefore allowing for much longer secure distance in the same fiber or free-space channels). The weak performance of antisqueezed-state-based CV QKD can be explained by the sensitivity of RR protocols to the noise in the state preparation \cite{Filip2008,Usenko2010a,Usenko2016}. On the contrary, in the case of DR the antisqueezed protocol can tolerate more channel loss than the squeezed-state one (demonstrating very poor results) and even outperforms the coherent-state protocol. In a telecom fiber with attenuation of $-0.2 dB/km$ the higher robustness of the antisqueezed DR UD CV QKD at the considered levels of noise would result in an almost-double increase in the maximum secure distance (from 4.5 to 7.5 km) compared to the coherent-state protocol. Note that the positive effect of antisqueezing noise of signal states is observed in the noisy channels and can be seen as the manifestation of the effect known as "fighting noise with noise," when noise on the reference side of the protocol makes it more robust against channel noise \cite{Garcia2009,Usenko2016}. In this regime, quantum squeezing of signal states in terms of the sub-shot-noise fluctuations may, in principle, not be needed and signal states with above-shot-noise fluctuations in the modulated ($x$-) quadrature and shot-noise and even above-shot-noise fluctuations in the unmodulated ($p$-) quadrature can be sufficient for improving the robustness of the DR UD CV QKD once impure signal states are considered.

For symmetrical channels and in the limit of infinitely strong modulation of pure squeezed states, the lower bound on the key rate in the DR scenario can be simplified as
\begin{multline}
K_{DR}\big|_{V_M\to\infty}=\\
=(\log_2{e})\big[C ArcTanh\frac{1}{C}-1\big]+\log_2{\frac{\eta|1-V_S|}{1+\eta|1-V_S|}},
\end{multline}
where $C\equiv\sqrt{\big[1+\eta(1/V_S-1)\big]\big[1+\eta(V_S-1)\big]}$. For the coherent-state protocol $V_S=1$ the expression further simplifies as 
\begin{equation}
K_{DR}^{(coh)}\big|_{V_M\to\infty}=\log_2{2\eta}-\frac{1}{2}\log_2[\eta(1-\eta)]-\log_2{e},
\end{equation}

which is lower by $\log_2{[e]}-1 \approx 0.44$ than the asymptotic expression for the lower bound on the key rate for the standard coherent-state protocol upon DR, being $\frac{1}{2}\log_2{\frac{\eta}{1-\eta}}$ \cite{Usenko2016}.

On the other hand, in the case of the RR scenario, in the limit of infinitely strong modulation the key rate in the symmetrical channel reads
\begin{multline}
K_{RR}\big|_{V_M\to\infty}=\\
=\frac{D}{2}\big[\log_2{\frac{D+1}{2}}-\log_2{\frac{D-1}{2}}\big]-\log_2{[1+\eta|1-V_S|]}-\\
-\log_2{e},
\end{multline}
where $D\equiv\sqrt{\frac{1+\eta(V_S-1)}{\eta V_S}}$. This can be further simplified for the coherent-state protocol, i.e., for $V_S=1$, as
\begin{equation}
K_{RR}^{(coh)}\big|_{V_M\to\infty}=\frac{1}{\ln{2}}\Bigg[\frac{ArcTanh(\sqrt{\eta})}{\sqrt{\eta}}-1\Bigg],
\end{equation}

which, in the limit of low transmittance $\eta\to 0$, can be well approximated by $\frac{\eta\log_2{e}}{3}$, being lower by a factor of $2/3$ than the similar limit for the standard coherent-state CV QKD protocol \cite{Usenko2015}.

\begin{figure}[h]
\begin{tabular}{ll}
\includegraphics[width=0.24\textwidth]{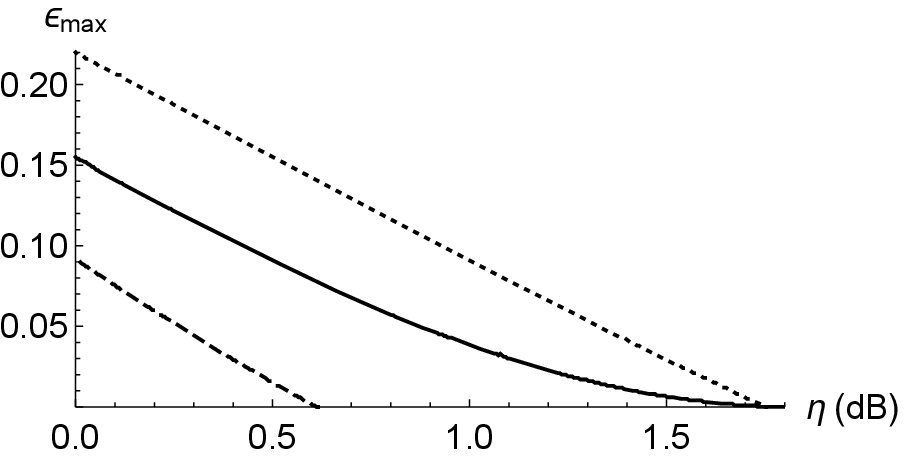}
\includegraphics[width=0.24\textwidth]{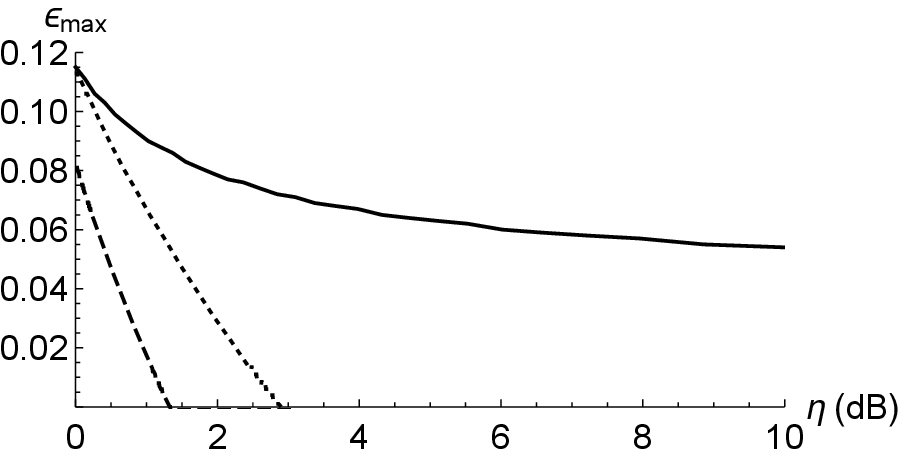}	
\end{tabular}
\caption{Maximal tolerable channel noise $\epsilon$ versus channel attenuation (on dB scale) for the protocols, secure against collective attacks for DR (left) and RR (right), based on coherent states (solid lines), squeezed states ($V_S=0.5$; dashed lines) and antisqueezed states ($V_S=2$; dotted lines). Modulation variance $V_M=100$. 
\label{emax}}
\end{figure}

We observe similar behavior (disadvantage of squeezing or antisqueezing in the RR scenario and advantage of antisqueezing in the DR scenario even compared to the coherent-state protocol) if we consider the robustness to excess channel noise at a given transmittance in the case of symmetrical channels, as plotted in Fig. \ref{emax}. 
Indeed, while the coherent-state protocol is more robust against channel noise in the case of RR, the antisqueezed protocol can tolerate larger amounts of channel excess noise once DR is used. It is evident from the plot that the antisqueezed protocol can tolerate about $50\%$ more channel noise than the coherent-state UD CV QKD. Therefore, surprisingly, modulation of a noisy antisqueezed quadrature (having more noise than the standard shot-noise level of a coherent state) can be advantageous for the UD CV QKD in the short-range quantum channels, increasing the key rate, the secure distance, and the tolerable channel noise of the protocol. 
\section{Summary and Conclusions}
\label{sec:sum}
We have considered the possibility of using squeezed or antisqueezed signal states in the unidimensional continuous-variable quantum key distribution protocol based on the Gaussian modulation of a single quadrature. The results show that squeezing or antisqueezing of the signal affects the physicality and security bounds of the protocol in the general case of phase-insensitive channels. In the typical case of phase-insensitive (symmetrical) channels the coherent-state protocol outperforms it's squeezed- and antisqueezed-state counterparts once reverse reconciliation is used. On the other hand, the antisqueezed-state protocol, based on the modulation of a quadrature, having more noise than a standard shot-noise level of a coherent state, demonstrates higher key rates and better robustness to losses and channel excess noise than coherent- and squeezed-state protocols. The result will be useful for the development of secure quantum communication systems upon short distances using a simplified single-quadrature modulation scheme and compact sources of squeezed light.
%
%
\acknowledgements
Author acknowledges support from the project LTC17086 of INTER-EXCELLENCE program of the Czech Ministry of Education and COST Action CA15220, 'QTSpace'.

\end{document}